\pdfoutput=1

\documentclass[11pt,twoside,a4paper,cmspaper,final,collab]{cms-tdr}
\def\svnVersion{203645}\def\cmsCernNo{2013-065}\def\cmsCernDate{\today}\def\cmsMessage{Published in the Journal of High Energy Physics as \href{http://dx.doi.org/10.1007/JHEP07(2013)163}{\doi{10.1007/JHEP07(2013)163.}}}

\begin{document}\cmsNoteHeader{BPH-11-013}

\hyphenation{had-ron-i-za-tion}
\hyphenation{cal-or-i-me-ter}
\hyphenation{de-vices}
\RCS$Revision: 189371 $
\RCS$HeadURL: svn+ssh://svn.cern.ch/reps/tdr2/papers/BPH-11-013/trunk/BPH-11-013.tex $
\RCS$Id: BPH-11-013.tex 189371 2013-06-07 22:26:29Z frmeier $
\providecommand{\re}{\ensuremath{\cmsSymbolFace{e}}}
\newcommand\Lb{\ensuremath{\Lambda_\mathrm{b}^0}} 
\newcommand\Lbbar{\ensuremath{\overline{\Lambda}_\mathrm{b}^0}} 
\providecommand\Ks{\PKzS}
\newcommand\pmu{\Pgm}
\newcommand\pmup{\Pgmp}
\newcommand\pmum{\Pgmm}
\newcommand\ppr{\Pp}
\newcommand\ppi{\Pgp}
\newcommand\ppip{\Pgpp}
\newcommand\ppim{\Pgpm}
\newcommand\bqrk{\cPqb\xspace}

\newcommand\chisq{\ensuremath{\chi^2}\xspace}

\renewcommand{\MeVc}{\MeV}
\renewcommand{\MeVcc}{\MeV}
\renewcommand{\GeVc}{\GeV}
\renewcommand{\GeVcc}{\GeV}
\renewcommand{\TeVc}{\TeV}
\renewcommand{\TeVcc}{\TeV}

\hyphenation{sig-ni-fi-cance}
\hyphenation{ATLAS}

\newcommand\vdef[1]{\expandafter\def\csname #1\endcsname}
\newcommand\vuse[1]{\csname #1\endcsname}

\cmsNoteHeader{BPH-11-013} 
\title{Measurement of the \Lb{} lifetime in pp collisions at $\sqrt{s}=7\TeV$}

\date{\today}

\abstract{
A measurement of the $\Lb$ lifetime using the decay $\Lb\to\JPsi\Lambda$ in proton-proton collisions at $\sqrt{s}=7\TeV$ is presented. The data set, corresponding to an integrated luminosity of about 5\fbinv, was recorded with the CMS experiment at the Large Hadron Collider using triggers that selected dimuon events in the $\JPsi$ mass region. The $\Lb$ lifetime is measured to be $1.503\pm0.052\stat\pm0.031\syst\unit{ps}.$
}

\hypersetup{%
pdfauthor={CMS Collaboration},%
pdftitle={Measurement of the Lambda(b)0 lifetime in pp collisions at sqrt(s) = 7 TeV},%
pdfsubject={CMS},%
pdfkeywords={CMS, physics}}

\maketitle 

\vdef{lumi:int}{\ensuremath{{5.08} } }
\vdef{cms:bfield}{3.8\unit{T}}

\vdef{pdg:L0:mass}{\ensuremath{1.116\GeVcc}}
\vdef{pdg:L0:ctau}{\ensuremath{7.89\unit{cm}}}
\vdef{pdg:L0:tau}{\ensuremath{263.1\pm\2.0\unit{ps}}}
\vdef{pdg:Lb:mass}{\ensuremath{5.620\GeVcc}}
\vdef{pdg:Lb:ctau}{\ensuremath{417\mum}}
\vdef{pdg:Lb:tau}{\ensuremath{1.39\unit{ps}}}
\vdef{pdg:Lb:tauE}{\ensuremath{1.39^{+0.038}_{-0.037}\unit{ps}}}
\vdef{pdg:Ks:mass}{\ensuremath{0.4976\GeVcc}}
\vdef{pdg:Ks:ctau}{\ensuremath{2.68\unit{cm}}}
\vdef{pdg:Ks:tau}{\ensuremath{89.6\pm\2.0\unit{ps}}}
\vdef{pdg:Jp:mass}{\ensuremath{3.097\GeVcc}}

\vdef{MELECTRON}{0.0005}
\vdef{MMUON}{0.1057}
\vdef{MPION}{0.1396}
\vdef{MKAON}{0.4937}
\vdef{MPROTON}{0.9383}
\vdef{MKSHORT}{0.4976}
\vdef{MKSTAR}{0.8960}
\vdef{MPHI}{1.0195}
\vdef{MLAMBDA_0}{1.1157}
\vdef{MD0}{1.8648}
\vdef{MDS}{1.9682}
\vdef{MDPLUS}{1.8693}
\vdef{MDSTARPLUS}{2.0100}
\vdef{MLAMBDA_C}{2.2865}
\vdef{MJPSI}{3.0969}
\vdef{MB_0}{5.2795}
\vdef{MBPLUS}{5.2792}
\vdef{MBS}{5.3663}
\vdef{MLAMBDA_B}{5.6202}

\vdef{D0:2012:Lb:lifetime:value}{1.303}
\vdef{D0:2012:Lb:lifetime:stat}{0.075}
\vdef{D0:2012:Lb:lifetime:syst}{0.035}
\vdef{ATLAS:2012:Lb:lifetime:value}{1.449}
\vdef{ATLAS:2012:Lb:lifetime:stat}{0.036}
\vdef{ATLAS:2012:Lb:lifetime:syst}{0.017}
\vdef{CDF:2011:Lb:lifetime:value}{1.537}
\vdef{CDF:2011:Lb:lifetime:stat}{0.045}
\vdef{CDF:2011:Lb:lifetime:syst}{0.014}

\vdef{sample:run2011:primdataset}{\texttt{MuOnia}}
\vdef{sample:run2011A1:name}{/MuOnia/Run2011A-May10ReReco\_v1/AOD}
\vdef{sample:run2011A1:evts}{16\,159\,072}
\vdef{sample:run2011A1:lumi}{234}
\vdef{sample:run2011A4:name}{/MuOnia/Run2011A-PromptReco\_v4/AOD}
\vdef{sample:run2011A4:evts}{24\,158\,927}
\vdef{sample:run2011A4:lumi}{892}
\vdef{sample:run2011A5:name}{/MuOnia/Run2011A-PromptReco\_v5/AOD}
\vdef{sample:run2011A5:evts}{6\,410\,953}
\vdef{sample:run2011A5:lumi}{434}
\vdef{sample:run2011A6:name}{/MuOnia/Run2011A-PromptReco\_v6/AOD}
\vdef{sample:run2011A6:evts}{9\,204\,408}
\vdef{sample:run2011A6:lumi}{722}
\vdef{sample:run2011B:name}{/MuOnia/Run2011B-PromptReco\_v1/AOD}
\vdef{sample:run2011B:evts}{>4\,084\,472}
\vdef{sample:run2011B:evts}{26\,767\,697}
\vdef{sample:run2011B:lumi}{2805}
\vdef{sample:run2011:evts}{82\,701\,057}
\vdef{sample:run2011:lumi}{5078}
\vdef{sample:run2011:json}{Cert\_160404-180252\_7TeV\_PromptReco\_Collisions11\_JSON\_MuonPhys.txt}

\vdef{sample:MC_Lb:name}{/LambdaBToJPsiMuMu\_2MuPEtaFilter\_7TeV-pythia6-evtgen/Fall10-START38\_V12-v2}
\vdef{sample:MC_Lb:shortname}{LambdaBToJPsiMuMu}
\vdef{sample:MC_Lb:evts}{2\,184\,854}
\vdef{sample:MC_Lb:feff}{0.000064}
\vdef{sample:MC_Lb:xsec}{8\,000\,000}
\vdef{sample:MC_Lb:lumi}{4267}

\vdef{sample:MC_B0:name}{/B0ToPsiMuMu\_2MuPEtaFilter\_Tight\_7TeV-pythia6-evtgen/Summer11-PU\_S4\_START42\_V11-v1/}
\vdef{sample:MC_B0:shortname}{B0ToPsiMuMu}
\vdef{sample:MC_B0:evts}{6\,012\,749}
\vdef{sample:MC_B0:feff}{0.000311}
\vdef{sample:MC_B0:xsec}{26\,500\,000}
\vdef{sample:MC_B0:lumi}{730}

\vdef{sample:MC_Bp:name}{/BpToPsiMuMu\_2MuPEtaFilter\_Tight\_7TeV-pythia6-evtgen/Summer11-PU\_S4\_START42\_V11-v1}
\vdef{sample:MC_Bp:shortname}{BpToJPsiMuMu}
\vdef{sample:MC_Bp:evts}{8\,389\,825}
\vdef{sample:MC_Bp:feff}{0.000319}
\vdef{sample:MC_Bp:xsec}{29\,800\,000}
\vdef{sample:MC_Bp:lumi}{883}

\vdef{sample:MC_Bs:name}{/BsToPsiMuMu\_2MuPEtaFilter\_Tight\_7TeV-pythia6-evtgen/Summer11-PU\_S4\_START42\_V11-v1}
\vdef{sample:MC_Bs:shortname}{BsToJPsiMuMu}
\vdef{sample:MC_Bs:evts}{1\,542\,913}
\vdef{sample:MC_Bs:feff}{0.0001}
\vdef{sample:MC_Bs:xsec}{24\,100\,000}
\vdef{sample:MC_Bs:lumi}{640}

\vdef{sample:MC_JpsiMuMu:name}{/JPsiToMuMu\_2MuPEtaFilter\_7TeV-pythia6-evtgen/Summer11-PU\_S4\_START42\_V11-v2}
\vdef{sample:MC_JpsiMuMu:shortname}{JPsiToMuMu}
\vdef{sample:MC_JpsiMuMu:evts}{38\,139\,484}
\vdef{sample:MC_JpsiMuMu:feff}{0.04923}
\vdef{sample:MC_JpsiMuMu:xsec}{12\,550\,000}
\vdef{sample:MC_JpsiMuMu:lumi}{96.5}

\vdef{sample:MCpriv:Lb:lifetime}{1.424\unit{ps}}
\vdef{sample:MCpriv:Lb:lifetime_ct}{427\mum}
\vdef{sample:MCpriv:Lb:evts}{2.47\,M}
\vdef{sample:MCpriv:Lb:NevtsCutBarrel}{3800}

\vdef{sample:MCpriv:B0:lifetime}{1.536\unit{ps}}
\vdef{sample:MCpriv:B0:lifetime_ct}{461\mum}
\vdef{sample:MCpriv:B0:evts}{22.0\,M}
\vdef{sample:MCpriv:B0:NevtsCutBarrel}{19'900}

\vdef{sample:MCoff:Lb:lifetime}{1.229\unit{ps}}

\vdef{sw:ver}{\url{CMSSW_4_2_8_patch7}}
\vdef{sw:pathBs2MuMu}{\url{HeavyFlavorAnalysis/Bs2MuMu}}
\vdef{sw:pathLb2JpL0}{\url{HeavyFlavorAnalysis/Lb2JpL0}}
\vdef{sw:pathlib}{\url{AnalysisDataFormats/HeavyFlavorObjects}}
\vdef{sw:pathUserCode}{\url{UserCode/frmeier/macrosLbLifetime}}
\vdef{sw:pathUserCodeMC}{\url{UserCode/frmeier/privateMCproduction}}
\vdef{sw:edanalyzer}{\texttt{HFLambdas}}
\vdef{sw:reader}{\texttt{lambdaReader}}
\vdef{sw:L0Effreader}{\texttt{lambdaEffReader}}
\vdef{sw:tag}{\texttt{--TODO:Tag--}}

\vdef{cuts:HLTbarrel:muMaxDoca}{\ensuremath{{<}0.5}}
\vdef{cuts:HLTbarrel:ptmumu}{\ensuremath{{>}6.5;10;13}}
\vdef{cuts:HLTbarrel:ymumu}{\ensuremath{{<}1.25}}
\vdef{cuts:HLTbarrel:mmumu}{\ensuremath{2.3\text{--}3.35}}
\vdef{cuts:HLTbarrel:probmumu}{\ensuremath{{>}0.5\%}}

\vdef{cuts:ntuple:verbose}{0}
\vdef{cuts:ntuple:tracksLabel}{generalTracks}
\vdef{cuts:ntuple:PrimaryVertexLabel}{offlinePrimaryVertices}
\vdef{cuts:ntuple:muonsLabel}{muons}
\vdef{cuts:ntuple:muonPt}{2.9}
\vdef{cuts:ntuple:muonType}{TrackerMuonArbitrated}
\vdef{cuts:ntuple:pionPt}{0.2}
\vdef{cuts:ntuple:protonPt}{0.9}
\vdef{cuts:ntuple:fTrackNormChi2}{7}
\vdef{cuts:ntuple:psiMuons}{2}
\vdef{cuts:ntuple:psiWindow}{0.3}
\vdef{cuts:ntuple:ksWindow}{0.08}
\vdef{cuts:ntuple:L0Window}{0.03}
\vdef{cuts:ntuple:LbWindow}{0.8}
\vdef{cuts:ntuple:B0Window}{0.8}
\vdef{cuts:ntuple:useAnalysisValuesForEff}{0}
\vdef{cuts:ntuple:PsiEffWindow}{0.015}
\vdef{cuts:ntuple:L0EffWindow}{0.12}
\vdef{cuts:ntuple:EffMaxChi2}{3.84}
\vdef{cuts:ntuple:EffMin3d}{0.5}
\vdef{cuts:ntuple:EffMaxDoca}{0.05}
\vdef{cuts:ntuple:deltaR}{99}
\vdef{cuts:ntuple:maxDoca}{0.5}
\vdef{cuts:ntuple:pAngle}{0.02}
\vdef{cuts:ntuple:maxVtxCut}{3.6}
\vdef{cuts:ntuple:useV0}{0}
\vdef{cuts:ntuple:doVcands}{0}
\vdef{cuts:ntuple:removeCandTracksFromVertex}{1}

\vdef{cuts:ntuple:verbose}{0}
\vdef{cuts:redtree:mjpWindow}{0.30}
\vdef{cuts:redtree:mjpWindow}{0.30}
\vdef{cuts:redtree:mrsWindow}{0.05}
\vdef{cuts:redtree:mrsWindow}{0.05}
\vdef{cuts:redtree:MuId1}{4}
\vdef{cuts:redtree:MuId1}{4}
\vdef{cuts:redtree:MuId2}{4}
\vdef{cuts:redtree:MuId2}{4}
\vdef{cuts:redtree:ptha1GTptha2}{1}
\vdef{cuts:redtree:ptha1Min}{0.0}
\vdef{cuts:redtree:ptha1Min}{0.0}
\vdef{cuts:redtree:ptha2Min}{0.0}
\vdef{cuts:redtree:ptha2Min}{0.0}
\vdef{cuts:redtree:ptjpMin}{0.0}
\vdef{cuts:redtree:ptjpMin}{0.0}
\vdef{cuts:redtree:ptmuMin}{2.9}
\vdef{cuts:redtree:ptmuMin}{2.99}
\vdef{cuts:redtree:ptrsMin}{0.0}
\vdef{cuts:redtree:ptrsMin}{0.0}

\vdef{cuts:analLb:rptmu1}{\ensuremath{> 3}}
\vdef{cuts:analLb:rptmu2}{\ensuremath{> 3}}
\vdef{cuts:analLb:mrs}{\ensuremath{0.006}}
\vdef{cuts:analLb:Kshypo}{\ensuremath{0.01}}
\vdef{cuts:analLb:rptha1}{\ensuremath{rptha1>rptha2}}
\vdef{cuts:analLb:rptha1}{\ensuremath{{>} 1.8}}
\vdef{cuts:analLb:rptha2}{\ensuremath{{>} 0.46}}
\vdef{cuts:analLb:probrs}{\ensuremath{{>} 0.1}}
\vdef{cuts:analLb:alphars}{\ensuremath{{<} 0.015}}
\vdef{cuts:analLb:d3rs/d3Ers}{\ensuremath{{>} 10}}
\vdef{cuts:analLb:vrrs}{\ensuremath{{>}3}}

\vdef{cuts:analB0:rptmu1}{\ensuremath{{>}3}}
\vdef{cuts:analB0:rptmu2}{\ensuremath{{>}3}}
\vdef{cuts:analB0:mrs}{\ensuremath{0.012}}
\vdef{cuts:analB0:L0hypo}{\ensuremath{0.01}}
\vdef{cuts:analB0:rptha1}{\ensuremath{{>}1.8}}
\vdef{cuts:analB0:rptha2}{\ensuremath{{>}0.5}}
\vdef{cuts:analB0:probrs}{\ensuremath{{>}0.15}}
\vdef{cuts:analB0:alphars}{\ensuremath{{<}0.012}}
\vdef{cuts:analB0:d3rs/d3Ers}{\ensuremath{{>}15}}
\vdef{cuts:analB0:vrrs}{\ensuremath{{>}3}}

\vdef{cuts:lambdaeff:truthmatchPrDeltaR}{\ensuremath{{<}0.1}}
\vdef{cuts:lambdaeff:truthmatchPiDeltaR}{\ensuremath{{<}0.2}}
\vdef{cuts:lambdaeff:truthmatchVtxRatio}{\ensuremath{{<}1.3}}
\vdef{cuts:lambdaeff:truthmatchVtxInvRatio}{\ensuremath{{>}0.77}}

\vdef{cuts:lambda:truthmatchPrDeltaR}{\ensuremath{{<}0.1}}
\vdef{cuts:lambda:truthmatchPiDeltaR}{\ensuremath{{<}0.2}}
\vdef{cuts:lambda:truthmatchVtxInvRatio}{\ensuremath{0.77}}
\vdef{cuts:lambda:truthmatchVtxRatio}{\ensuremath{1.3}}

\vdef{reader:lambda:HLTmatchDeltaR}{0.5}

\vdef{fit:cutoff:B0:masslo}{5.16}
\vdef{fit:cutoff:B0:masshi}{5.75}
\vdef{fit:cutoff:B0:BsRange}{5.4}
\vdef{fit:cutoff:Lb:masslo}{5.40}
\vdef{fit:cutoff:Lb:masshi}{6.00}
\vdef{fit:cutoff:tlo}{-1}
\vdef{fit:cutoff:thi}{15}

\vdef{result:B0:data:Nevt:value}{6772}
\vdef{result:B0:data:Nevt:stat}{87}
\vdef{result:B0:data:mass:value}{5278.9}
\vdef{result:B0:data:mass:stat}{0.2}
\vdef{result:B0:data:lifetime:value}{1.526}
\vdef{result:B0:data:lifetime:stat}{0.019}
\vdef{result:B0:data:lifetime:syst}{0.017}

\vdef{result:Lb:data:Nevt:value}{1013}
\vdef{result:Lb:data:Nevt:stat}{40}
\vdef{result:Lb:data:mass:value}{5619.7}
\vdef{result:Lb:data:mass:stat}{0.5}
\vdef{result:Lb:data:lifetime:value}{1.503}
\vdef{result:Lb:data:lifetime:stat}{0.052}
\vdef{result:Lb:data:lifetime:syst}{0.031}

\vdef{result:ratio:data:lifetime:value}{0.98}
\vdef{result:ratio:data:lifetime:stat}{0.04}
\vdef{result:ratio:data:lifetime:syst}{0.01}

\vdef{systerr:}{\ensuremath{}}
\vdef{systerr:}{\ensuremath{}}
\vdef{systerr:}{\ensuremath{}}
\vdef{systerr:}{\ensuremath{}}
\vdef{systerr:}{\ensuremath{}}

\vdef{MCclosure:B0:all}{\ensuremath{1.536\pm0.001\unit{ps}}}
\vdef{MCclosure:B0:candmatch}{\ensuremath{1.534\pm0.004\unit{ps}}}
\vdef{MCclosure:B0:anacut}{\ensuremath{1.534\pm0.007\unit{ps}}}
\vdef{MCclosure:B0:anacuttrig}{\ensuremath{1.530\pm0.012\unit{ps}}}

\vdef{MCclosure:LbLbbar:all}{\ensuremath{1.423\pm0.001\unit{ps}}}
\vdef{MCclosure:LbLbbar:candmatch}{\ensuremath{1.406\pm0.009\unit{ps}}}
\vdef{MCclosure:LbLbbar:anacut}{\ensuremath{1.424\pm0.012\unit{ps}}}
\vdef{MCclosure:LbLbbar:anacuttrig}{\ensuremath{1.430\pm0.022\unit{ps}}}

\vdef{MCclosure:Lb:all}{\ensuremath{1.424\pm0.002\unit{ps}}}
\vdef{MCclosure:Lb:candmatch}{\ensuremath{1.411\pm0.012\unit{ps}}}
\vdef{MCclosure:Lb:anacut}{\ensuremath{1.429\pm0.017\unit{ps}}}
\vdef{MCclosure:Lb:anacuttrig}{\ensuremath{1.436\pm0.030\unit{ps}}}

\vdef{MCclosure:Lbbar:all}{\ensuremath{1.423\pm0.002\unit{ps}}}
\vdef{MCclosure:Lbbar:candmatch}{\ensuremath{1.401\pm0.012\unit{ps}}}
\vdef{MCclosure:Lbbar:anacut}{\ensuremath{1.417\pm0.018\unit{ps}}}
\vdef{MCclosure:Lbbar:anacuttrig}{\ensuremath{1.423\pm0.031\unit{ps}}}

\vdef{reso:sigma_B0_truthmatched_ct}{\ensuremath{0.101\pm0.000\unit{ps}}}
\vdef{reso:sigma_B0_truthmatched_d3}{\ensuremath{138\pm0.4\mum}}
\vdef{reso:sigma_B0_truthmatched_p}{\ensuremath{0.073\pm0.000\GeVc}}
\vdef{reso:sigma_B0_anacut_ct}{\ensuremath{0.081\pm0.001\unit{ps}}}
\vdef{reso:sigma_B0_anacut_d3}{\ensuremath{114\pm0.8\mum}}
\vdef{reso:sigma_B0_anacut_p}{\ensuremath{0.066\pm0.001\GeVc}}

\vdef{reso:bias_B0_truthmatched_ct}{\ensuremath{+0.001\pm0.000\unit{ps}}}
\vdef{reso:bias_B0_truthmatched_d3}{\ensuremath{+2.4\pm0.3\mum}}
\vdef{reso:bias_B0_truthmatched_p}{\ensuremath{0.000\pm0.000\GeVc}}
\vdef{reso:bias_B0_anacut_ct}{\ensuremath{0.000\pm0.001\unit{ps}}}
\vdef{reso:bias_B0_anacut_d3}{\ensuremath{+0.6\pm0.9\mum}}
\vdef{reso:bias_B0_anacut_p}{\ensuremath{0.000\pm0.001\GeVc}}

\vdef{reso:sigma_lb_truthmatched_ct}{\ensuremath{0.096\pm0.001\unit{ps}}}
\vdef{reso:sigma_lb_truthmatched_d3}{\ensuremath{137\pm1\mum}}
\vdef{reso:sigma_lb_truthmatched_p}{\ensuremath{0.111\pm0.001\GeVc}}
\vdef{reso:sigma_lb_anacut_ct}{\ensuremath{0.084\pm0.001\unit{ps}}}
\vdef{reso:sigma_lb_anacut_d3}{\ensuremath{113\pm2\mum}}
\vdef{reso:sigma_lb_anacut_p}{\ensuremath{0.081\pm0.002\GeVc}}

\vdef{reso:bias_lb_truthmatched_ct}{\ensuremath{0.000\pm0.001\unit{ps}}}
\vdef{reso:bias_lb_truthmatched_d3}{\ensuremath{+0.5\pm0.9\mum}}
\vdef{reso:bias_lb_truthmatched_p}{\ensuremath{+0.004\pm0.001\GeVc}}
\vdef{reso:bias_lb_anacut_ct}{\ensuremath{-0.001\pm0.001\unit{ps}}}
\vdef{reso:bias_lb_anacut_d3}{\ensuremath{+0.4\pm1.7\mum}}
\vdef{reso:bias_lb_anacut_p}{\ensuremath{+0.004\pm0.001\GeVc}}

\vdef{MCclosureFit:B0}{\ensuremath{1.532\pm0.010\unit{ps}}}
\vdef{MCclosureFit:Lb}{\ensuremath{1.430\pm0.021\unit{ps}}}
\vdef{MCclosureFit:LbLb}{\ensuremath{1.428\pm0.029\unit{ps}}}
\vdef{MCclosureFit:LbLbbar}{\ensuremath{1.432\pm0.031\unit{ps}}}

\vdef{MCclosureFit:deviation:B0}{\ensuremath{-0.004\unit{ps}}}
\vdef{MCclosureFit:deviation:Lb}{\ensuremath{+0.006\unit{ps}}}
\vdef{MCclosureFit:deviation:LbLb}{\ensuremath{+0.004\unit{ps}}}
\vdef{MCclosureFit:deviation:LbLbbar}{\ensuremath{+0.008\unit{ps}}}

\section{Introduction}
The study of \bqrk~baryons is a necessary ingredient to understand \bqrk-hadron phenomenology. The heavy-quark expansion model of nonperturbative quantum chromodynamics provides a framework for predicting properties of heavy-flavour hadrons, including their lifetimes.  Here, a simple description is used where the heavy \bqrk~quark is surrounded by a light quark or diquark system.  Estimates can be made of the lifetime and of the ratio of lifetimes between particles sharing the same heavy-quark flavour~\cite{Shifman:1986mx,Bigi:1992su,Bigi:1997fj,Beneke:2002rj,Franco:2002fc,Tarantino:2003qw,Tarantino:2005zi,Lenz:2008xt}.  The early calculation predicted a spread of the lifetimes of order 5\% among all b hadrons~\cite{Shifman:1986mx} and the ratio of the \Lb{} and \PBz{} lifetimes, $\tau_{\Lb{}}/\tau_{\PBz{}}$, to be greater than 0.90~\cite{arXiv1207.1158}. The initial measurements of the \bqrk-baryon lifetime were generally lower than predicted~\cite{arXiv1207.1158}, which motivated more refined calculations. This resulted in predictions as low as $\tau_{\Lb{}}/\tau_{\PBz{}} = 0.86 \pm 0.05$~\cite{Gabbiani:2004tp}. An overview of the current state of the predictions and measurements can be found in Ref.~\cite{arXiv1207.1158}.  Measurements of the \Lb{} lifetime prior to 2011 can be found in Refs.~\cite{Buskulic:1992nu,Abreu:1993vm,Akers:1993sa,Abreu:1995me,Akers:1995hp,Buskulic:1995xt,Abreu:1996nt,Akers:1995ui,Abe:1996df,Ackerstaff:1997qi,Barate:1997if,Abreu:1999hu,Abazov:2004bn,Abazov:2007sf,Abazov:2007al,Abulencia:2006dr,Aaltonen:2009zn}. More recent measurements of the \Lb{} lifetime include: $\vuse{CDF:2011:Lb:lifetime:value} \pm \vuse{CDF:2011:Lb:lifetime:stat} \pm \vuse{CDF:2011:Lb:lifetime:syst}$\unit{ps} from CDF~\cite{CDFLb}, $\vuse{D0:2012:Lb:lifetime:value} \pm \vuse{D0:2012:Lb:lifetime:stat} \pm \vuse{D0:2012:Lb:lifetime:syst}$\unit{ps} from D0~\cite{D0Lb}, and $\vuse{ATLAS:2012:Lb:lifetime:value} \pm \vuse{ATLAS:2012:Lb:lifetime:stat} \pm \vuse{ATLAS:2012:Lb:lifetime:syst}$\unit{ps} from ATLAS~\cite{ATLASLb}, where the first uncertainties are statistical and the second are systematic.

In this paper, a measurement of $\tau_{\Lb}$ is presented, using the decay $\Lb\rightarrow \JPsi\PgL$, with $\PgL\rightarrow \ppr\ppim$ and $\JPsi\rightarrow \pmup\pmum$. The kinematically analogous channel $\PBz\rightarrow\JPsi\Ks$, with $\Ks\rightarrow\ppip\ppim$, is used as a cross-check, with selection criteria similar to those for the \Lb{} analysis. Charge-conjugate states are assumed throughout this paper.  The measurement is made using proton-proton collision data at $\sqrt{s}=7$\TeV recorded by the Compact Muon Solenoid (CMS) experiment operating at the Large Hadron Collider (LHC). The data set for this measurement was collected in 2011 using \JPsi{}-enriched dimuon triggers, and corresponds to an integrated luminosity of about 5\fbinv~\cite{CMSlumi}.

\section{CMS detector}
The central feature of the CMS apparatus is a superconducting solenoid of 6\unit{m} internal diameter. The main subdetectors used for the analysis are the silicon tracker, consisting of silicon pixel and silicon strip layers, and the muon system.   The tracker is immersed in a 3.8\unit{T} axial magnetic field of the superconducting solenoid. The pixel tracker consists of three barrel layers and two endcap disks at each barrel end.   The strip tracker has 10 barrel layers and 12 endcap disks at each barrel end. The tracker provides an impact parameter resolution of ${\sim}15\mum$ and a transverse momentum (\pt) resolution of about 1.5\% for 100\GeV particles. Charged hadrons, including pions and protons, are not explicitly identified by their type.  Muons are measured in gas-ionisation detectors that are embedded in the steel return yoke outside the solenoid.   In the barrel, there is a drift tube system interspersed with resistive plate chambers, and in the endcaps there is a cathode strip chamber system, also interspersed with resistive plate chambers.  The first-level trigger used in this analysis is based on the muon system alone, while the high-level trigger uses additional information from the tracker.  A detailed description of the CMS detector can be found in Ref.~\cite{JINST}.

The CMS experiment uses a right-handed coordinate system, with the origin at the nominal interaction point, the $x$ axis pointing towards the centre of the LHC ring, the $y$ axis pointing up (perpendicular to the plane of the LHC ring), and the $z$ axis along the anticlockwise-beam direction.   The polar angle $\theta$ is measured from the positive $z$ axis and the pseudorapidity is defined by $\eta=-\ln[\tan(\theta/2)]$.   The azimuthal angle is measured from the positive $x$ axis in the plane perpendicular to the beam.

\section{Event selection and efficiency modelling}
Dimuon triggers optimised for selecting events with \JPsi{} candidates are used.  The trigger requires two oppositely charged muons with an invariant mass compatible with the \JPsi{}-meson mass.  The dimuon candidate must also be found in the central region (dimuon rapidity $\abs{y_{\pmup\pmum}}<1.25$),  which is the region with the best impact parameter and dimuon invariant-mass resolution. With increasing instantaneous luminosity, the trigger requirements were adjusted several times during the data-taking period.  In the course of data taking, the dimuon mass window was changed from 2.5--4.0\GeVcc{} to 2.30--3.35\GeVcc{}, while the minimum transverse momentum of the dimuon candidate was increased from 6.5\GeVc{} to 13\GeVc{}.  Additional requirements were also added during this period:  the distance of closest approach between the two muons was required to be less than 0.5\unit{cm}, the vertex-fit \chisq{} probability of the two muons was required to be greater than 0.5\%, and the two muons were required to bend away from each other in the tracker.

The four charged particles (\pmup\pmum\ppr\ppim) in the decay channel $\Lb\rightarrow\JPsi\PgL$ allow for a full reconstruction of the \Lb{} baryon. The selection requirements are chosen to maximise the ratio of signal yield to the square root of the signal-plus-background yield.  Events with two oppositely charged muons are selected. The muons are reconstructed using information from the tracker and the muon system. The muon candidates must be within the kinematic acceptance of the detector by demanding the muon transverse momentum $\pt^{\pmu}$ satisfies $\pt^{\pmu}>3.5$\GeV for muon pseudorapidity $\abs{\eta^\mu}<1.2$, and $\pt^{\pmu}>3.0$--3.5\GeV for $1.2<\abs{\eta^{\pmu}}<1.6$, where the $\pt^{\pmu}$ threshold decreases linearly as a function of $\abs{\eta^\mu}$.  The muon candidates are also required to have a track $\chi^2$ per degree of freedom less than 1.8, at least 11 tracker hits, at least two hits in the pixel system, a match to at least one track segment in the muon system, and a transverse (longitudinal) impact parameter less than 3\unit{cm} (30\unit{cm}) with respect to the primary event vertex.   The preliminary choice for the primary vertex is the one with the highest sum of the squares of $\pt$ of the tracks associated with it. The two muons are then used to form a \JPsi{} candidate.

A \PgL{} candidate is formed using oppositely charged tracks with the proton candidate required to have $\pt\vuse{cuts:analLb:rptha1}\GeVc{}$ and the pion candidate $\pt\vuse{cuts:analLb:rptha2}$\GeVc{}. The higher-momentum track is taken as the proton.

Kinematic vertex fits are used to identify the \Lb{} candidates~\cite{KinVertFit}.  An initial unconstrained fit is used to measure three selection parameters. First, the proton-pion invariant mass ($m_{\ppr\pi^-}$) is found from the \PgL{} candidate tracks and is required to be within 6\MeVcc{} of the world-average \PgL{} mass~\cite{PDG}.  The \PgL{} candidates are rejected if the dipion invariant mass $m_{\ppip\ppim}$ is within 10\MeVcc{} of the \Ks{} mass~\cite{PDG}, when the proton candidate is assigned a pion mass.  Then, the dimuon invariant mass ($m_{\mu^+\mu^-}$) is determined from an unconstrained fit to the \JPsi{} candidate tracks and is required to be within 300\MeVcc{} of its world-average value~\cite{PDG}.  Finally, a pointing angle is measured for the \PgL{} candidate, which is defined as the angle between the momentum of the reconstructed \PgL{} particle and the vector between its production and decay vertices.  The pointing angle is required to be less than 0.015\unit{rad}.

To determine the proper decay time~$t$ of a \Lb{} candidate, another kinematic vertex fit is performed where the two muon tracks and the \PgL{} candidate are constrained to come from a common vertex, with the \PgL{} and \JPsi{} candidate masses constrained to their respective nominal values~\cite{PDG}.  The \PgL{} vertex-fit \chisq{} probability must be greater than 10\%.  The separation between the \PgL{} vertex and the eventual primary vertex (as defined below) must be larger than 10$\sigma$, where $\sigma$ is the calculated uncertainty in the relative position.  In addition, the \PgL{} vertex must be at least 3\unit{cm} away from the mean pp collision position in the transverse plane.

Multiple pp interactions per bunch crossing are present in the data. The reconstructed event vertex with the smallest impact parameter in the $z$ direction to the \Lb{} candidate trajectory is selected as the primary production vertex.  The position of the primary vertex is recalculated excluding the tracks from the \Lb{} decay if they were used for the initial primary-vertex reconstruction. The proper decay time~$t$ is found for each \Lb{} candidate by calculating the ratio of the decay length and the momentum of the \Lb{}, divided by the world-average \Lb{} mass~\cite{PDG}.

As a cross-check, the \PBz{} lifetime, $\tau_{\PBz}$, is determined using the $\PBz\rightarrow \JPsi\Ks$ channel.  A completely analogous procedure is followed to find the \PBz{} candidates, replacing the proton-track hypothesis with a pion-track hypothesis and fitting for a \Ks{} candidate instead of a \PgL{} candidate.  The \Ks{} candidates are formed assuming both tracks are pions (the higher-momentum one is required to have $\pt\vuse{cuts:analB0:rptha1}$\GeVc{}, the lower-momentum one to have $\pt\vuse{cuts:analB0:rptha2}$\GeVc), with $m_{\ppip\ppim}$ within 12\MeVcc{} of the world-average value~\cite{PDG}.  A candidate is vetoed if replacing the pion-mass hypothesis by the proton mass yields an invariant mass $m_{\ppr\ppim}$ within 10\MeVcc{} of the \PgL{} mass~\cite{PDG}.  The pointing angle and other kinematic requirements are the same as those for the \Lb{} selection.

\begin{figure}[hbtp]
    \begin{center}
	\includegraphics[width=.9\textwidth]{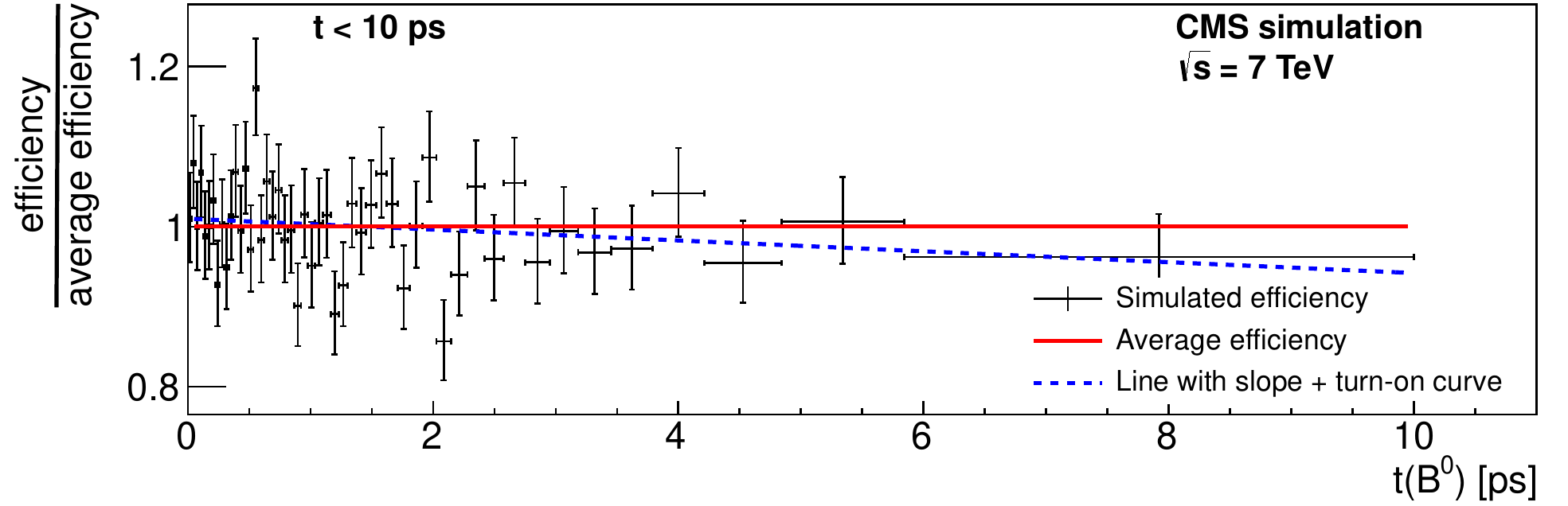}
	\includegraphics[width=.9\textwidth]{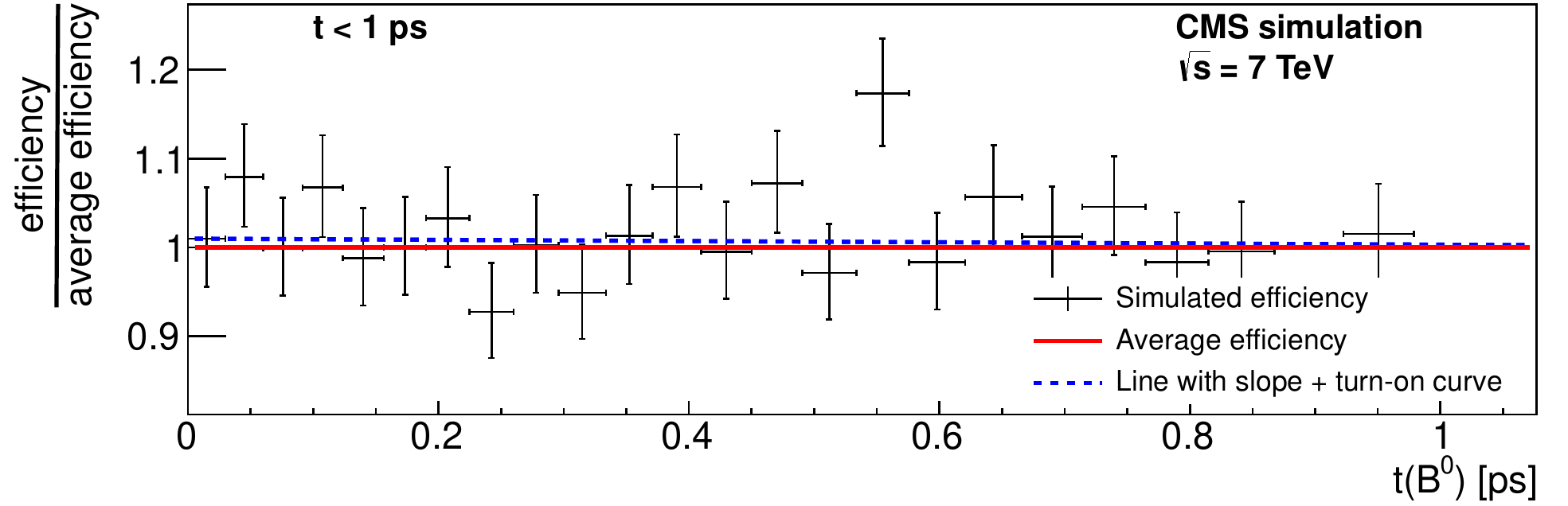}
	\includegraphics[width=.9\textwidth]{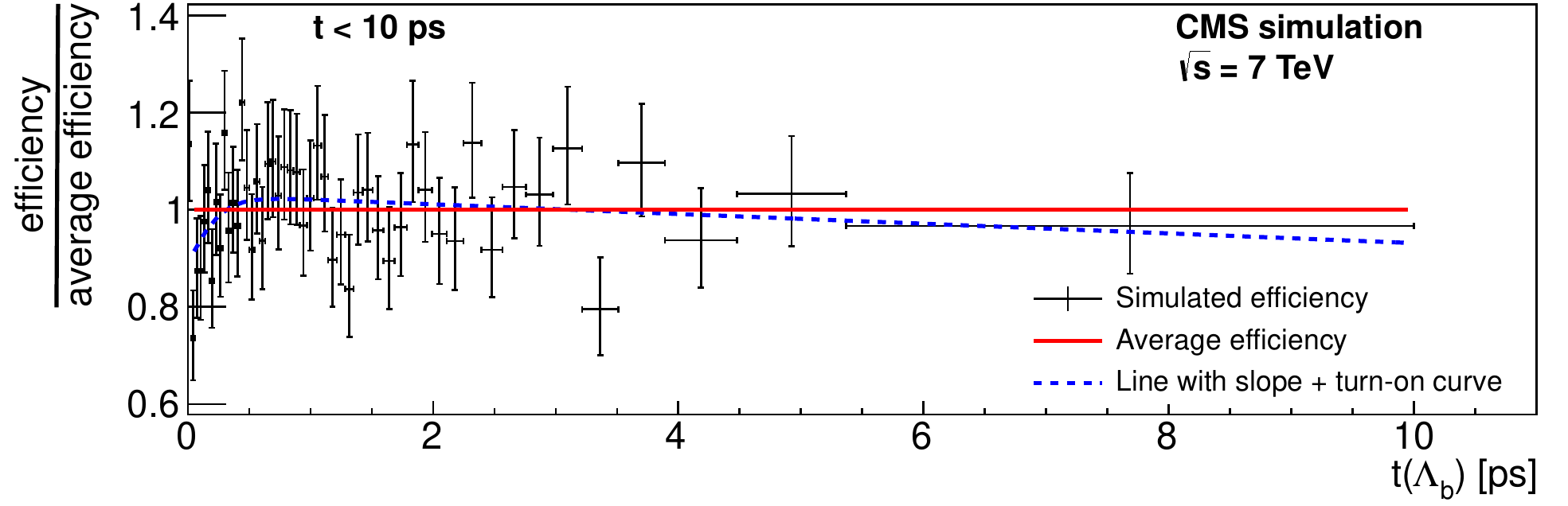}
	\includegraphics[width=.9\textwidth]{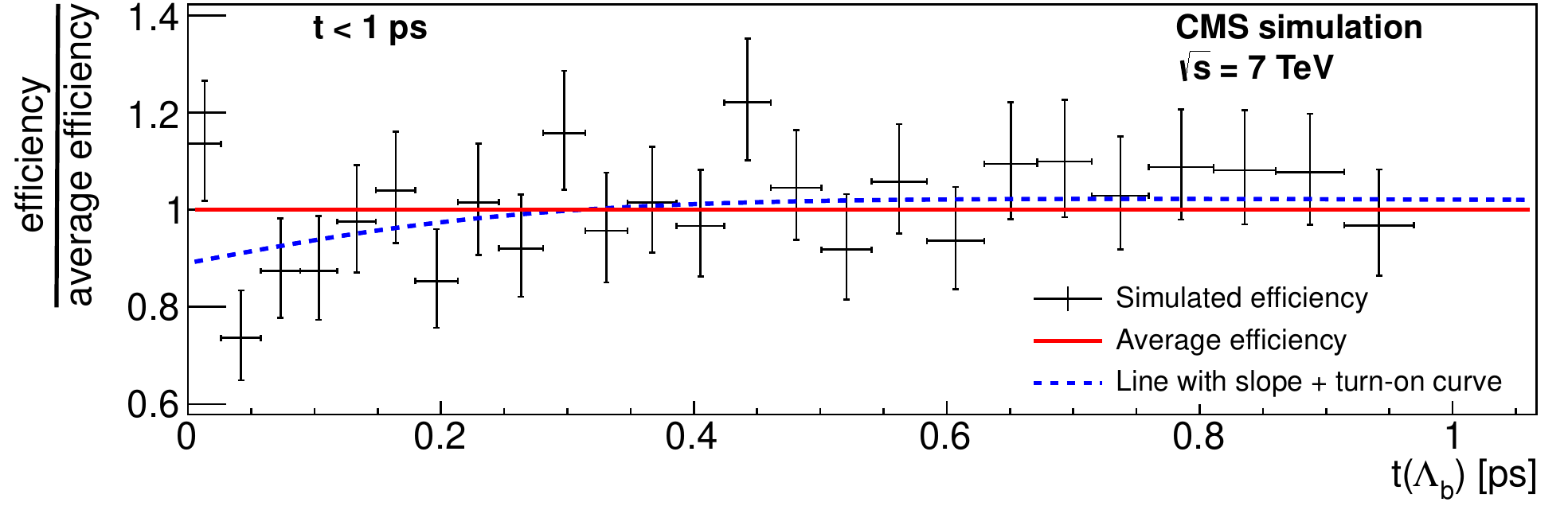}
	\caption{The ratio of the reconstruction and selection efficiencies to the overall average efficiency as a function of proper decay time for \PBz{} (upper two plots) and \Lb{} (lower two plots). The second and fourth plots show the \PBz{} and \Lb{} efficiency ratios, respectively, for smaller proper decay times $t<1\unit{ps}$. The horizontal solid lines show a ratio of 1, while the dashed lines display the results from fitting the efficiencies to the function given by Eq.~\eqref{eqn.efffunc}.}
	\label{fig.eff}
    \end{center}
\end{figure}

Simulated event samples are used to model the signal and background distributions.   The \PYTHIA 6.422~\cite{PYTHIA} event generator is used, with the \Lb{} lifetime fixed to 1.425\unit{ps}~\cite{PDG} and the \bqrk-hadron decays described by the \EVTGEN 9.1 simulation package~\cite{EVTGEN}.  The particle propagation and detector simulation is performed using \GEANTfour 9.4~\cite{GEANT4}, and the events are fully reconstructed with the same software as the data.  The differences between the reconstructed and generated values for the proper decay time, the flight length, and the candidate \Lb{} momentum are found to be compatible with zero in the simulated sample.  The sideband-subtracted data distributions in each of the selection variables are compared to those in simulated data and found to be consistent within their uncertainties.   The overall reconstruction and selection efficiency as a function of the proper decay time~$t$ is determined from the simulated signal samples by calculating the ratio of the numbers of reconstructed and generated \Lb{} (\PBz{}) candidates in bins of proper decay time.  Figure~\ref{fig.eff} displays the ratio of the reconstruction and selection efficiencies measured from simulation to the overall average efficiency, as a function of the proper decay time for the \PBz{} (top) and \Lb{} (bottom). The efficiencies are consistent with being independent of the proper decay time, as shown by the solid horizontal line in each plot.  The \Lb{} efficiency depends on various kinematic variables such as the \Lb{} transverse momentum.  Therefore, a comprehensive comparison of the distributions of these variables between the data and the Monte Carlo simulation was performed, and in all cases the distributions were found to be consistent.

\section{Proper decay time fit}
Unbinned extended maximum-likelihood fits are performed to determine the \Lb{} and \PBz{} lifetimes. The input variables are the invariant mass $m$, proper decay time~$t$, and its uncertainty $\sigma_t$, calculated per candidate from the kinematic vertex fit using full error propagation.  The likelihood fit is implemented using the \textsc{RooFit} 3.53 package~\cite{RooFit}.  The likelihood function (ignoring the normalisation terms for simplicity) is
\begin{linenomath*}
\begin{equation}\begin{split}
\label{eqn.fitfunc}
    \mathcal{L}  =  \prod_i \Big[ & N_\text{sig}         \cdot G2(m_i; m_\text{sig}, \sigma_{m_1}, \sigma_{m_2}, f) \cdot \re^{-t/\tau_\text{sig}} \otimes G(t_i;\mu, S\cdot\sigma_{t,i})  \\
	                &+ N_\text{prompt}     \cdot P(m_i; a)                                      \cdot G(t_i;\mu,S\cdot\sigma_{t,i}) \\
		        &+ N_\text{nonprompt} \cdot P(m_i; a)                                      \cdot \re^{-t/\tau_\text{nonprompt}} \otimes G(t_i;\mu,S\cdot\sigma_{t,i})  \Big] ,
\end{split}\end{equation}
\end{linenomath*}
where the index $i$ goes over the events, $N_\text{sig}$ is the number of signal events, $N_{\text{prompt}}$ is the number of prompt background events not coming from \bqrk-hadron decays, and $N_{\text{nonprompt}}$ is the corresponding number of nonprompt background events coming from \bqrk-hadron decays.
The prompt background is dominated by $\JPsi{}$ mesons directly produced in the pp collision, while the nonprompt background is dominated by $\JPsi{}$ mesons from decays of b hadrons. In both cases, the $\JPsi{}$ mesons are combined with real or misidentified $\PgL{}$ candidates from the event.  The parameters $\tau_\text{sig}$ and $\tau_{\text{nonprompt}}$ denote the lifetime of the signal and of nonprompt background, respectively. In Eq. \eqref{eqn.fitfunc}, the $G2$ function is the sum of two Gaussians with a common mean $m_\text{sig}$ and widths $\sigma_{m_1}$ and $\sigma_{m_2}$, and the parameter $f$ denotes the relative fraction of the area of the two Gaussians. The function $G$ refers to a single Gaussian describing the detector lifetime resolution, with a mean $\mu$ and a width $S\cdot\sigma_{t,i}$, where $S$ is a scale factor determined from the fit. This resolution function is common to all three likelihood components since the $\sigma_t$ distributions do not differ significantly.  The effect of using this simplifying assumption on $\sigma_t$ is evaluated as a systematic uncertainty and found to be negligible.  For the background components, the invariant-mass $m$ distribution is parameterised by a normalised first-degree polynomial of slope $a$, $P(m; a)$. The prompt and nonprompt backgrounds share the same slope. The maximum-likelihood fit to the data is performed allowing all parameters to vary.

\section{Results}
\begin{figure}[hbtp]
    \begin{center}
	\includegraphics[width=.9\textwidth]{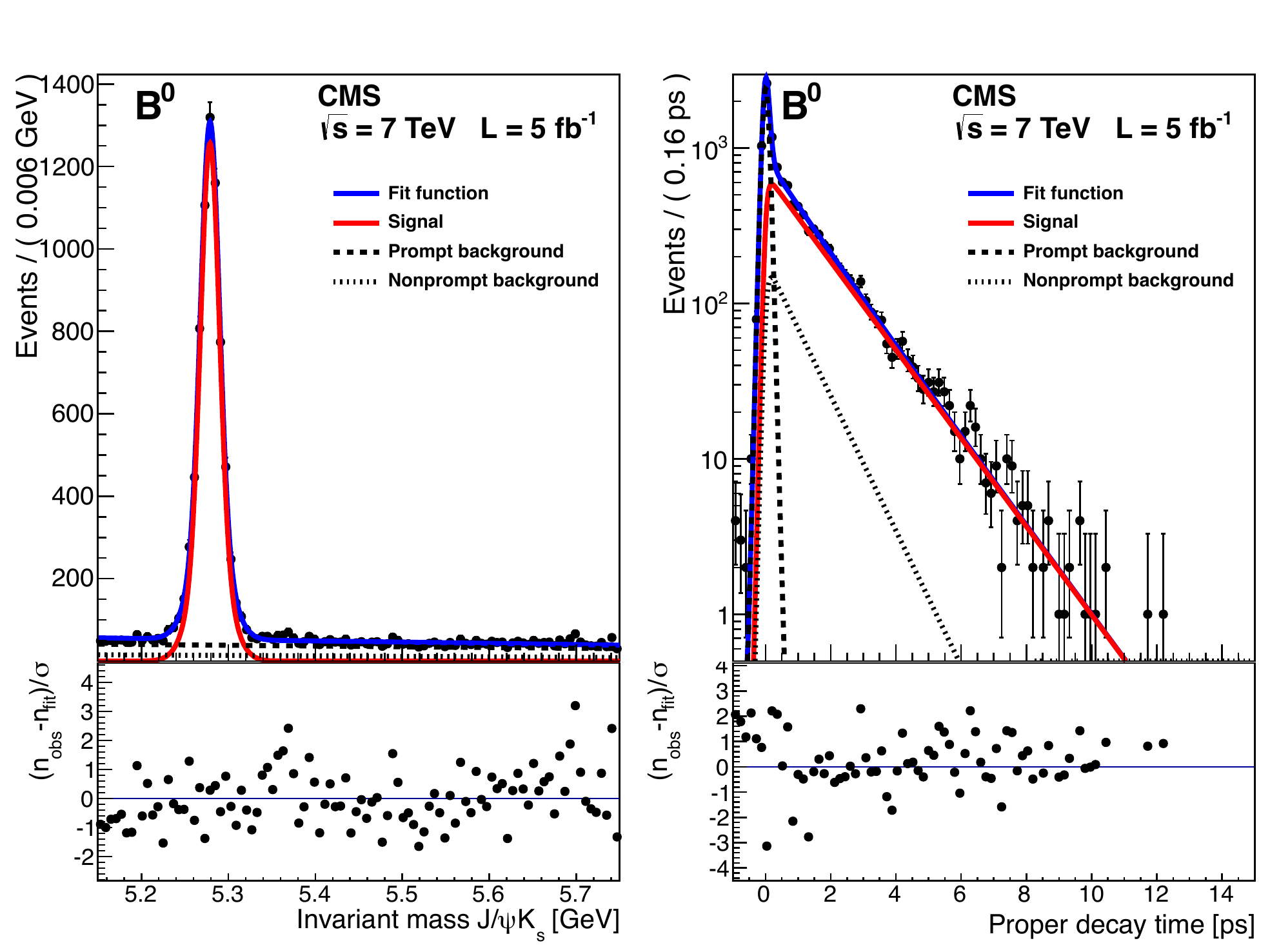}
	\includegraphics[width=.9\textwidth]{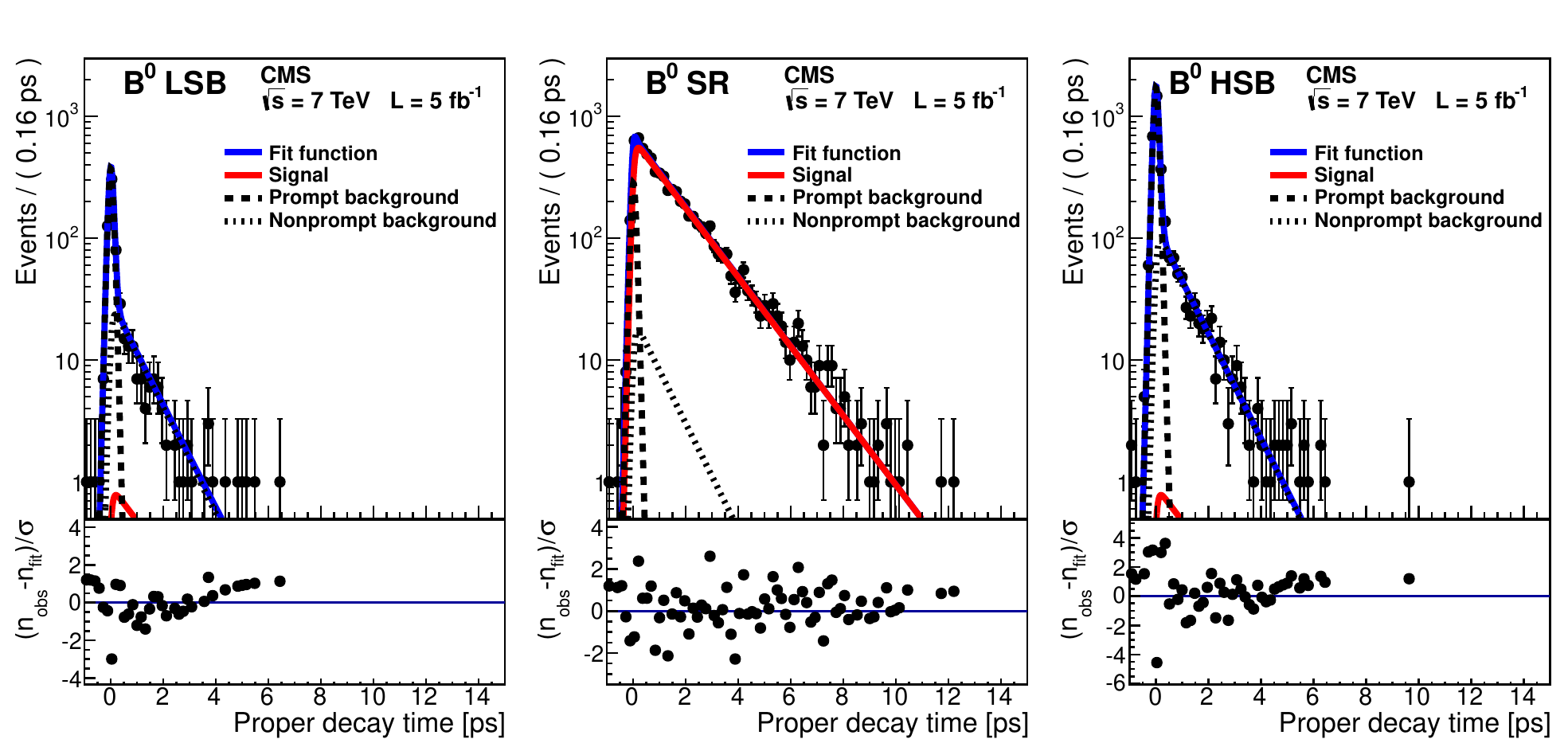}
	\caption{Projections of the invariant-mass and proper decay time distributions and the results of the fit are shown for the \PBz{} decay in the upper panels. The dark solid lines give the results of the overall fit to the data. The lighter solid lines are the signal contributions, and the dashed and dotted lines show the prompt and nonprompt background contributions, respectively. The lower panels display the proper decay time projections for the low-mass sideband (LSB, left), the signal (SR), and the high-mass sideband (HSB, right) regions defined in the text. The lower plots in each panel give the corresponding pull distributions for the data and fit results shown. All plots are from the same fit.}
	\label{fig.b0fit}
    \end{center}
\end{figure}

\begin{figure}[hbtp]
    \begin{center}
	\includegraphics[width=.9\textwidth]{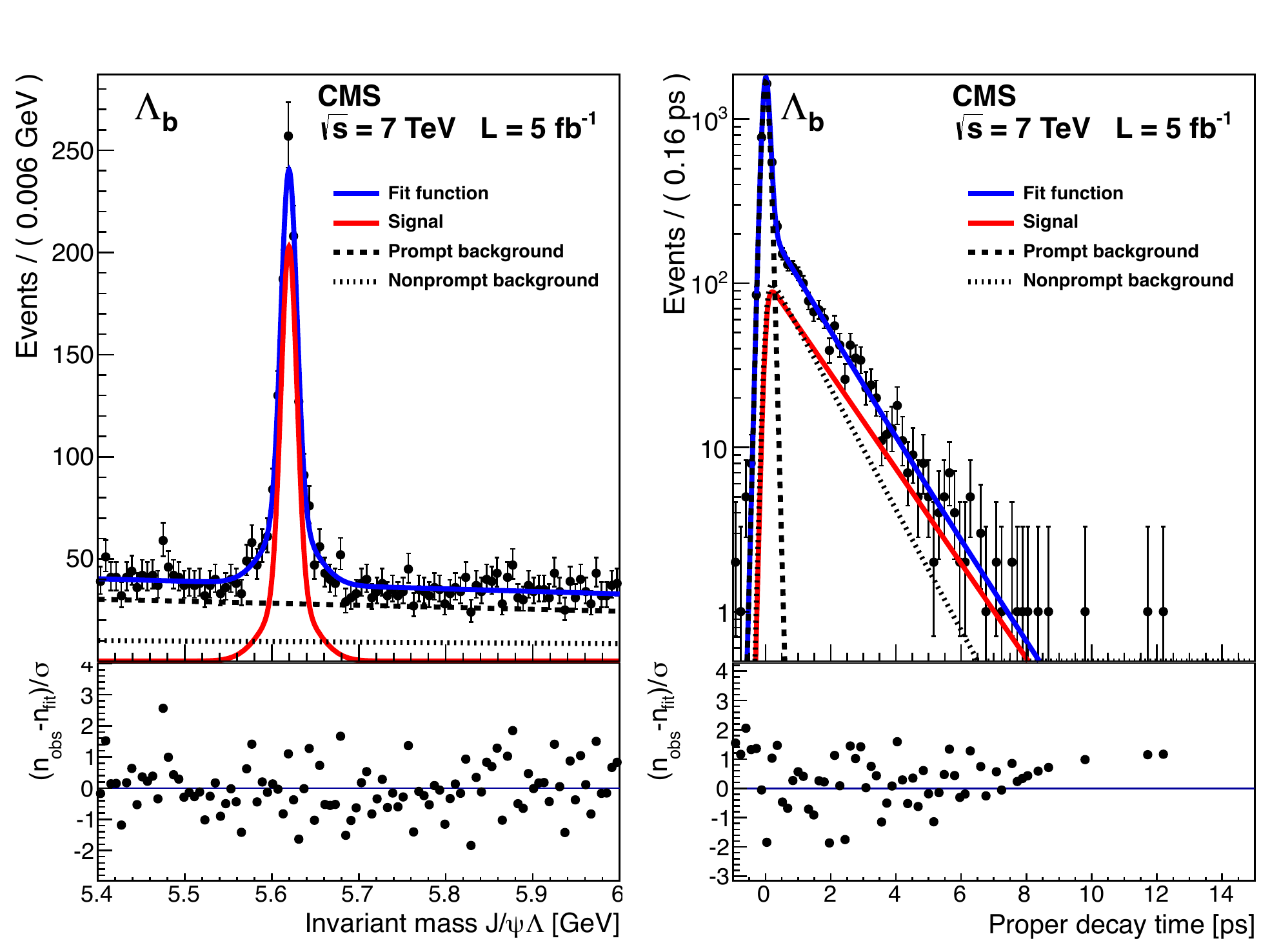}
	\includegraphics[width=.9\textwidth]{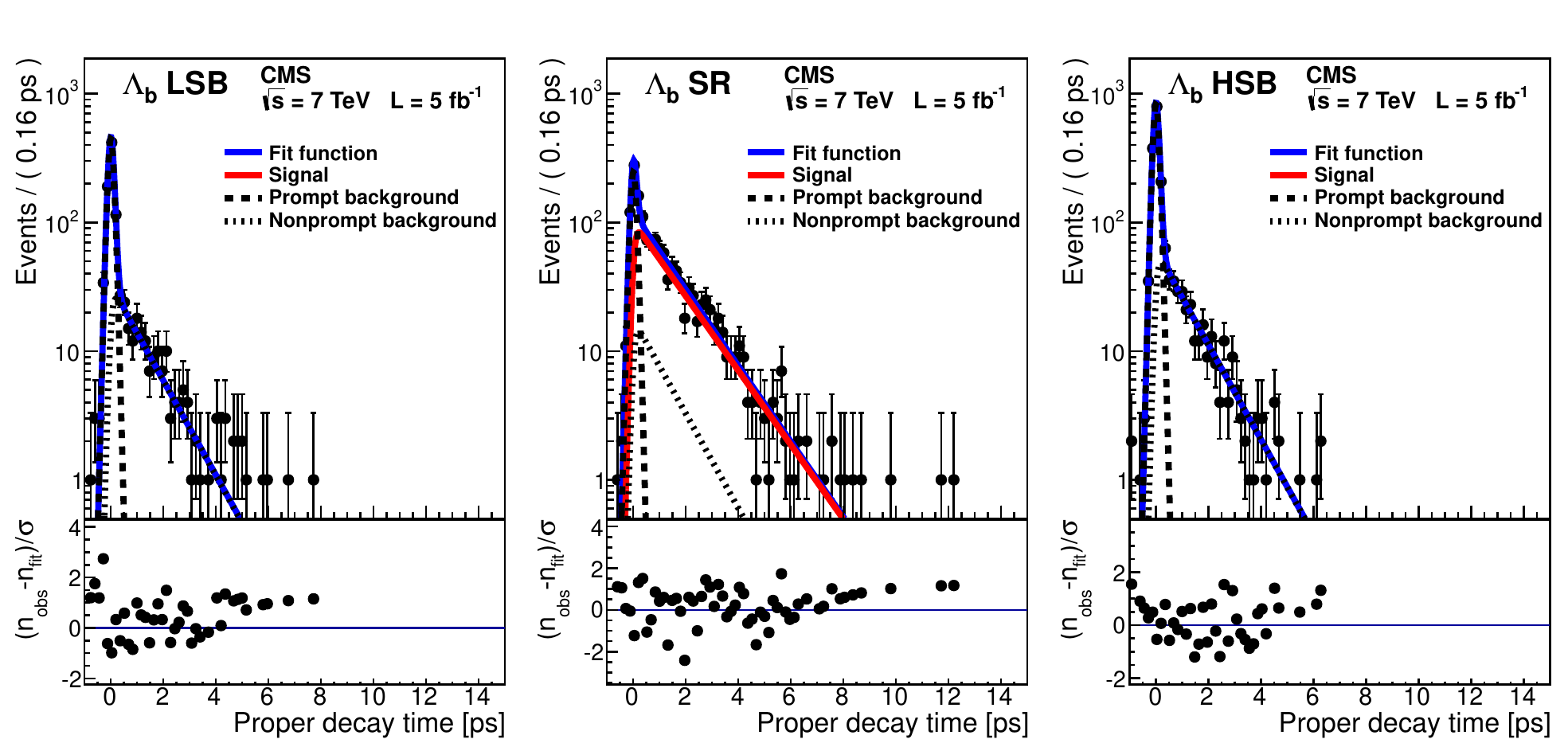}
	\caption{Projections of the invariant-mass and proper decay time distributions and the results of the fit are shown for the \Lb{} decay in the upper panels. The dark solid lines give the results of the overall fit to the data. The lighter solid lines are the signal contributions, and the dashed and dotted lines show the prompt and nonprompt background contributions, respectively. The lower panels display the proper decay time projections for the low-mass sideband (LSB, left), the signal (SR), and the high-mass sideband (HSB, right) regions defined in the text. The lower plots in each panel give the corresponding pull distributions for the data and fit results shown. All plots are from the same fit.}
	\label{fig.lbfit}
    \end{center}
\end{figure}

Projections of the invariant-mass and proper decay time distributions and the results of the fits for the \PBz{} and \Lb{} are shown in the upper panels of Figs.~\ref{fig.b0fit} and~\ref{fig.lbfit}, respectively. The lower panels in each figure give the proper decay time projections and fit results for low-mass sideband (left), signal (center), and high-mass sideband (right) regions of the invariant-mass distribution. The signal region is defined to be within 2$\sigma$ of the mass peak, where $\sigma$ is the mass resolution obtained by integrating the double-Gaussian signal function with its parameter values determined from the fit. The low-mass sideband region goes from 5.15 (5.4)\GeV to within 3$\sigma$ below the peak, and the high-mass sideband region runs from 3$\sigma$ above the peak to 5.75 (6.0)\GeV for the \PBz{} (\Lb{}). The lower plot in each panel of Figs.~\ref{fig.b0fit} and~\ref{fig.lbfit} displays the pull distribution for the corresponding data and fit results shown in the upper plot.

The results from the fits are summarised in Table~\ref{tab.results}, where the uncertainties are statistical only. The measured \PBz{} lifetime shown in the table is consistent with the world-average value of $1.519 \pm 0.007$\unit{ps}~\cite{PDG}. The mass values for both \Lb{} and \PBz{} given in the table compare well with the world-average values~\cite{PDG}.

The fitting procedure is validated by studying simulated pseudo-experiments in which the proper decay time distributions are generated using different lifetime values. The resulting lifetime measurements found from the fit are compatible with being unbiased, and the width of the pull distribution, $(\text{measured value} - \text{input value})/\text{uncertainty}$, is consistent with 1.0.

 \begin{table}[!htbp]
    \topcaption{Summary of the fit results for \Lb{} and \PBz{} with their statistical uncertainties. }
    \begin{center}
    \label{tab.results}
	\begin{tabular}{lccc}
	    \hline
	Hadron & $N_\text{sig}$ & $m$ (\MeVns{}) & $\tau$ (ps)\\
	    \hline
	\Lb{}  & $\vuse{result:Lb:data:Nevt:value}\pm\vuse{result:Lb:data:Nevt:stat}$ & $\vuse{result:Lb:data:mass:value}\pm\vuse{result:Lb:data:mass:stat}$ & $\vuse{result:Lb:data:lifetime:value}\pm\vuse{result:Lb:data:lifetime:stat}$ \\
	\PBz{}  & $\vuse{result:B0:data:Nevt:value}\pm\vuse{result:B0:data:Nevt:stat}$ & $\vuse{result:B0:data:mass:value}\pm\vuse{result:B0:data:mass:stat}$ & $\vuse{result:B0:data:lifetime:value}\pm\vuse{result:B0:data:lifetime:stat}$ \\
	    \hline
	\end{tabular}
    \end{center}
\end{table}

Sources of systematic uncertainty are detector alignment, efficiency as a function of proper decay time, event selection, and fit model.  To estimate possible effects due to uncertainties in the alignment, nine different simulated samples with distorted geometries are produced and analysed~\cite{Chatrchyan:2009sr}.   The lifetime difference between the nominal result and the sample that produces the largest deviation (scaled to the estimated residual misalignment present in the detector) is taken as the systematic uncertainty from this source.

Since the overall efficiency, determined through simulation, is consistent with being independent of the proper decay time, as shown in Fig.~\ref{fig.eff}, no efficiency correction is used in the lifetime result. Nevertheless, the effect of a possible proper-decay-time-dependent efficiency is included as a systematic uncertainty.
To this end, the efficiency is included in the likelihood function in Eq.~\eqref{eqn.fitfunc} as a function of the measured proper decay time. The difference between the lifetime found using a constant efficiency (\ie no efficiency in the likelihood) and that found using the fitted efficiency function is taken as a systematic uncertainty.  The efficiency is fit to the function
\begin{linenomath*} 
\begin{equation} \label{eqn.efffunc}
    \varepsilon(t) = p_0 \cdot \left(1+p_1t+\frac{p_2}{1+\re^{-t/p_3}}\right),
\end{equation}
\end{linenomath*}
where the free parameters $p_i$ are determined from the fit.  This parameterisation has the useful feature that as the parameter $p_2$ goes to 0, the function becomes a straight line, which is consistent with the behaviour of the \PBz{} efficiencies shown in Fig.~\ref{fig.eff}.  The results of fitting this function to the \PBz{} and \Lb{} efficiencies are shown by the dashed lines in Fig.~\ref{fig.eff}.

To account for possible biases from the event selection criteria, a systematic uncertainty is calculated from the difference between the observed and expected values in simulated events using the full analysis chain.

Simulated pseudo-experiments produced with different input parameter values and different modelling of the fit functions are used to estimate the corresponding systematic uncertainties on the lifetime measurement.  Variations include:   using two different lifetimes to describe the nonprompt background to control the possible presence of a second nonprompt background component, varying the prompt and nonprompt background, varying the lifetime of the nonprompt background to control possible correlation to the lifetime of the background, and using larger per-event uncertainties for the background components to control a possible mismodelling of the resolution.  Extending the likelihood function in Eq.~\eqref{eqn.fitfunc} from a single detector lifetime resolution $\sigma_t$ for all three components to individual resolutions per component showed a negligible effect. Table~\ref{tab.sys} gives the systematic uncertainties from the four sources and their sum in quadrature, which is taken as the overall systematic uncertainty on the \Lb{} lifetime measurement.

\begin{table}[!htbp]
    \topcaption{Summary of the systematic uncertainties in the \Lb{} lifetime measurement.}
    \begin{center}
    \label{tab.sys}
	\begin{tabular}{lccc}
	    \hline
	Source                 & Systematic uncertainty (ps) \\
	    \hline
	Alignment              & 0.005 \\
	Efficiency             & 0.030 \\
	Event selection        & 0.005 \\
	Fit model              & 0.004 \\
	\hline
	Total  & 0.031  \\
	    \hline
	\end{tabular}
    \end{center}
\end{table}

Checks are also performed to see if any detector effect leads to a systematic deviation not covered by the ones previously discussed.  This is done by dividing the data sample into parts, where each of the partitions is expected to give the same results.  The checks are performed with azimuthal angle, pseudorapidity, transverse momentum, run era, muons bending away or towards each other, and number of primary interaction vertices.   No statistically significant effects are seen.  For the \Lb{} channel, we also remove the \Ks{}-mass-veto requirement on the \PgL{} candidate and find a negligible effect on the lifetime result.

\section{Summary}

\begin{figure}[hbtp]
    \begin{center}
	\includegraphics[width=.9\textwidth]{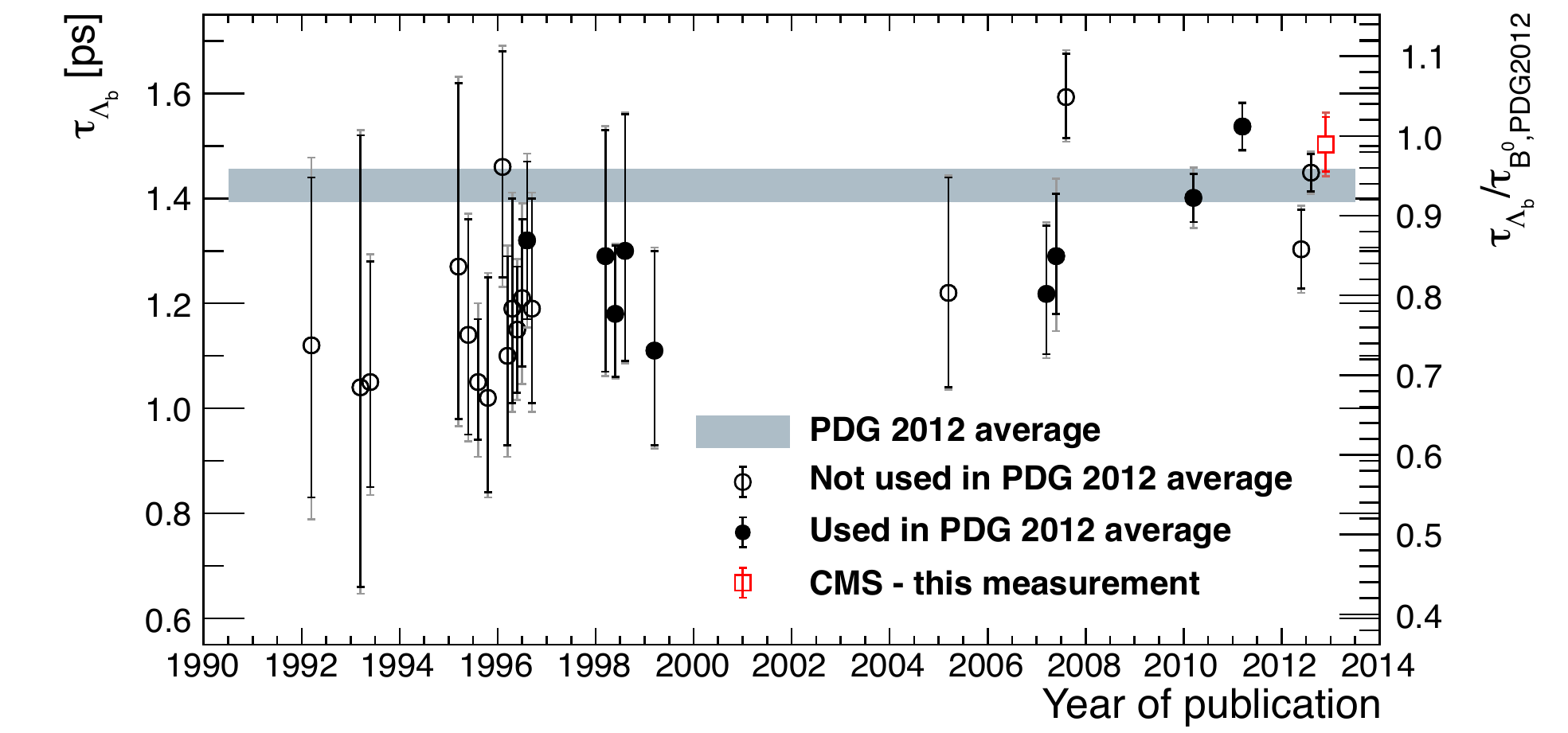}
	\caption{Evolution over time of the \Lb{} lifetime measurements (left scale)~\cite{Buskulic:1992nu,Abreu:1993vm,Akers:1993sa,Abreu:1995me,Akers:1995hp,Buskulic:1995xt,Abreu:1996nt,Akers:1995ui,Abe:1996df,Ackerstaff:1997qi,Barate:1997if,Abreu:1999hu,Abazov:2004bn,Abazov:2007sf,Abazov:2007al,Abulencia:2006dr,Aaltonen:2009zn,CDFLb,D0Lb,ATLASLb} and the ratio of the measurements to the 2012 world-average \PBz{} lifetime $\tau_\PBz{}$ \cite{PDG} (right scale). The values shown as open circles were not included in the Particle Data Group (PDG) 2012 average~\cite{PDG}, displayed as the grey band, while those shown as filled circles were included. The result of this analysis is shown by the open square.  The inner error bars represent the statistical uncertainties, and the outer error bars show the combined statistical and systematic uncertainties added in quadrature.  Where needed, points have been shifted slightly along the time axis to enhance clarity.}
	\label{fig.LbLifetimeEvolutionCMS}
    \end{center}
\end{figure}

A measurement of the \Lb{} lifetime has been presented using the decay \Lb{}$\rightarrow$\JPsi{}\PgL{} in pp collisions at $\sqrt{s} = 7$\TeV with the CMS detector.  From a data set corresponding to an integrated luminosity of about 5\fbinv{}, the \Lb{} lifetime is found to be $\tau_{\Lb{}} = \vuse{result:Lb:data:lifetime:value}\pm\vuse{result:Lb:data:lifetime:stat}\stat\pm\vuse{result:Lb:data:lifetime:syst}\syst\unit{ps}$. The kinematically similar decay \PBz{}$\rightarrow$\JPsi{}\Ks{} was used as a cross-check, confirming that no efficiency correction was needed.  The \Lb{} lifetime result is in agreement with the world-average value of $1.425 \pm 0.032$\unit{ps}~\cite{PDG} and has a precision comparable to that of other recent measurements~\cite{CDFLb,D0Lb,ATLASLb}.  As illustrated in Fig.~\ref{fig.LbLifetimeEvolutionCMS}, this new result confirms the tendency of the more recent measurements that give larger lifetimes, in better agreement with the early theoretical predictions~\cite{Shifman:1986mx,arXiv1207.1158}.

\section*{Acknowledgements}
We congratulate our colleagues in the CERN accelerator departments for the excellent performance of the LHC and thank the technical and administrative staffs at CERN and at other CMS institutes for their contributions to the success of the CMS effort. In addition, we gratefully acknowledge the computing centres and personnel of the Worldwide LHC Computing Grid for delivering so effectively the computing infrastructure essential to our analyses. Finally, we acknowledge the enduring support for the construction and operation of the LHC and the CMS detector provided by the following funding agencies: BMWF and FWF (Austria); FNRS and FWO (Belgium); CNPq, CAPES, FAPERJ, and FAPESP (Brazil); MEYS (Bulgaria); CERN; CAS, MoST, and NSFC (China); COLCIENCIAS (Colombia); MSES (Croatia); RPF (Cyprus); MoER, SF0690030s09 and ERDF (Estonia); Academy of Finland, MEC, and HIP (Finland); CEA and CNRS/IN2P3 (France); BMBF, DFG, and HGF (Germany); GSRT (Greece); OTKA and NKTH (Hungary); DAE and DST (India); IPM (Iran); SFI (Ireland); INFN (Italy); NRF and WCU (Republic of Korea); LAS (Lithuania); CINVESTAV, CONACYT, SEP, and UASLP-FAI (Mexico); MSI (New Zealand); PAEC (Pakistan); MSHE and NSC (Poland); FCT (Portugal); JINR (Armenia, Belarus, Georgia, Ukraine, Uzbekistan); MON, RosAtom, RAS and RFBR (Russia); MSTD (Serbia); SEIDI and CPAN (Spain); Swiss Funding Agencies (Switzerland); NSC (Taipei); ThEPCenter, IPST and NSTDA (Thailand); TUBITAK and TAEK (Turkey); NASU (Ukraine); STFC (United Kingdom); DOE and NSF (USA).

Individuals have received support from the Marie-Curie programme and the European Research Council and EPLANET (European Union); the Leventis Foundation; the A. P. Sloan Foundation; the Alexander von Humboldt Foundation; the Belgian Federal Science Policy Office; the Fonds pour la Formation \`a la Recherche dans l'Industrie et dans l'Agriculture (FRIA-Belgium); the Agentschap voor Innovatie door Wetenschap en Technologie (IWT-Belgium); the Ministry of Education, Youth and Sports (MEYS) of Czech Republic; the Council of Science and Industrial Research, India; the Compagnia di San Paolo (Torino); the HOMING PLUS programme of Foundation for Polish Science, cofinanced by EU, Regional Development Fund; and the Thalis and Aristeia programmes cofinanced by EU-ESF and the Greek NSRF.

\bibliography{auto_generated}   

\cleardoublepage \appendix\section{The CMS Collaboration \label{app:collab}}\begin{sloppypar}\hyphenpenalty=5000\widowpenalty=500\clubpenalty=5000\textbf{Yerevan Physics Institute,  Yerevan,  Armenia}\\*[0pt]
S.~Chatrchyan, V.~Khachatryan, A.M.~Sirunyan, A.~Tumasyan
\vskip\cmsinstskip
\textbf{Institut f\"{u}r Hochenergiephysik der OeAW,  Wien,  Austria}\\*[0pt]
W.~Adam, T.~Bergauer, M.~Dragicevic, J.~Er\"{o}, C.~Fabjan\cmsAuthorMark{1}, M.~Friedl, R.~Fr\"{u}hwirth\cmsAuthorMark{1}, V.M.~Ghete, N.~H\"{o}rmann, J.~Hrubec, M.~Jeitler\cmsAuthorMark{1}, W.~Kiesenhofer, V.~Kn\"{u}nz, M.~Krammer\cmsAuthorMark{1}, I.~Kr\"{a}tschmer, D.~Liko, I.~Mikulec, D.~Rabady\cmsAuthorMark{2}, B.~Rahbaran, C.~Rohringer, H.~Rohringer, R.~Sch\"{o}fbeck, J.~Strauss, A.~Taurok, W.~Treberer-Treberspurg, W.~Waltenberger, C.-E.~Wulz\cmsAuthorMark{1}
% ...
\vskip\cmsinstskip
% ...
1:~~Also at Vienna University of Technology, Vienna, Austria\\
2:~~Also at CERN, European Organization for Nuclear Research, Geneva, Switzerland\\
% ...

\end{sloppypar}
\end{document}